\documentclass{emulateapj}
\usepackage{times}
\usepackage{amsmath}

\def\la{\mathrel{\mathpalette\fun <}}

\def\fun#1#2{\lower0.837ex\vbox{\baselineskip0ex\lineskip0.209ex
  \ialign{$\mathsurround=0ex#1\hfil##\hfil$\crcr#2\crcr\sim\crcr}}}

\def\msun{M_\odot}
\def\msunyr{M_\odot \ {\rm yr}^{-1}}

\def\sles{\lower2pt\hbox{$\buildrel {\scriptstyle <}
   \over {\scriptstyle\sim}$}}

\def\sgreat{\lower2pt\hbox{$\buildrel {\scriptstyle >}
   \over {\scriptstyle\sim}$}}

\def\la{\mathrel{\mathpalette\fun <}}

\def\aprop{\mathrel{\mathpalette\fun \propto}}

\begin{document}

\title{Constraining the Physics of AM Canum Venaticorum Systems with the Accretion
     Disk Instability Model}
\shortauthors{Cannizzo \& Nelemans}
\author{ 
         John~K.~Cannizzo\altaffilmark{1,2},
           Gijs~Nelemans\altaffilmark{3,4}
}
\altaffiltext{1}{CRESST and Astroparticle Physics
      Laboratory, NASA/GSFC, Greenbelt, MD 20771, USA;
                 John.K.Cannizzo@nasa.gov}
\altaffiltext{2}{Department of Physics, University of 
       Maryland, Baltimore County, 1000 Hilltop Circle,
           Baltimore, MD 21250, USA} 
\altaffiltext{3}{Department of Astrophysics/IMAPP, Radboud University Nijmegen,
            P.O. Box 9010, 6500 GL Nijmegen, The Netherlands}
\altaffiltext{4}{Institute for Astronomy, KU Leuven, Celestijnenlaan 200D, 
                3001 Leuven, Belgium}

\begin{abstract}
Recent work by Levitan et al has 
        expanded the long--term photometric
    database for AM CVn stars. In particular, 
their outburst properties are well--correlated with
orbital period, and allow constraints to be placed
   on the secular mass transfer rate between secondary
  and primary if one adopts the disk instability model
   for the outbursts. We use the observed range
   of outbursting behavior for AM CVn systems as a function
  of orbital period to place a constraint on 
mass transfer rate versus orbital period.
          We infer a rate $\sim$$5\times10^{-9}\msunyr(P_{\rm orb}/
   1000 \ {\rm s})^{-5.2}$.
  We show the 
   functional form so obtained is consistent with
 the recurrence time--orbital period relation found
by Levitan et al using a simple theory for the recurrence time. 
    Also, we predict their steep dependence of outburst duration on orbital
 period will flatten considerably once the longer orbital period
 systems  have more complete observations. 
\end{abstract}


\section{Introduction}

Cataclysmic variables (CVs) are
semi--detached binaries
   consisting of interacting stars in which a Roche--lobe
filling secondary transfers matter to a more massive, 
 and also more compact primary. 
     The dwarf novae (DNe) constitute
a subclass further characterized
     by semiregular outbursts of several magnitudes,
  recurring on timescale of days to years (Warner 1995a).
 The novalikes reside at higher mass transfer rates and do not
show DN outbursts.
     The limit cycle accretion disk instability model (DIM)
   has been successful in explaining the outbursts
  (Smak 1984).
  CVs evolve to shorter orbital periods, driven
 by angular momentum loss (AML) from a combination of
   a magnetic wind from the secondary star
  and gravitational radiation (Knigge, Baraffe, \& Patterson
   2011 $=$ KBP).
    Although the DIM is generally agreed to be the
  correct explanation for DN outbursts, mainly
  because the observed dividing line between steady
  and outbursting  systems (i.e., DNe and novalikes)
     agrees with theory (Smak 1983b),
       further progress has come slowly.
  For instance, any theory for the recurrence time for DN
  outbursts (Cannizzo, Shafter, Wheeler 1988 $=$ CSW)
    is muddied by the fact that the outburst
     properties at a given orbital period
                                   exhibit a wide scatter
    (Warner 1995a).

The AM CVn stars are a subset of CVs at very short
orbital period, less than an 
 hour, whose spectra are dominated by helium (Nelemans 2005, Solheim 2010).
   The prototype, AM CVn, was discovered neary fifty years ago
  (Smak 1967). Paczy\'nski  (1967) 
proposed the system to be a short orbital period binary with two
degenerate, He-–rich stars with a period evolution driven by
gravitational wave radiation (GWR), but he considered detached rather
than semi-–detached binaries.
   Motivated by observational inferences of mass transfer
  in AM CVn, 
 namely rapid photometric flickering 
 (Warner \& Robinson 1972), 
               Faulkner et al. (1972) 
  presented 
      the first self--consistent 
  model for AM CVns, 
    correctly  taking into account
                   their  semi--detached nature.

       The coincidence in stable 
   periodic variations in spectral lines
     with photometric variations can sometimes
             be used to obtain orbital 
periods, and the weak X--ray emission indicates a white dwarf (WD)
  primary  accretor rather than a neutron star  (Nather et al. 1981, 
         Patterson et al. 1993, Groot et al. 2001).  
 AM CVn stars are thought to have evolved beyond the minimum 
orbital period which divides nondegenerate and degenerate 
secondaries. The systems span a wide range in mass transfer
  rate ${\dot M}_T$.  In fact, in the AM CVn stars one sees
   a range spanning not only the high to intermediate range
   equivalent to the novalike $\rightarrow$ DNe transition 
          in normal hydrogen--containing
CVs, but also a range spanning intermediate to low ${\dot M}_T$
values.  For the low ${\dot M}_T$ systems the disk is too cool
   to have DIM outbursts.

In some sense the AM CVn stars are a better laboratory
for the DIM than normal DNe because less scatter is expected in 
their  ${\dot M}_T$ values at a given orbital  period,
    given that GWR is the sole AML mechanism.  
    KBP find that the normal (i.e., solar composition)
     short orbital period
 DNe below the period gap must have AML enhanced by a factor
  $\sim$2.47 above that given solely by GWR in order to produce
the observed period gap.  This only applies to an ensemble
  average; as with the DNe above the gap, the outburst properties
   of systems below the gap show considerable scatter at a given
orbital  period.
             In addition, ${\dot M}_T$
 varies by more than  four orders of magnitude across the range of  orbital  periods
  in the AM CVn systems, which makes orbital period--dependent
   properties of the outbursts
     more noticeable.

Amassing a large database for AM CVn stars has
been hampered by their faintness. Now, thanks 
to the results of dedicated transient surveys, an
   avalanche of new systems and data on outbursting behavior
   has become available. 
   Levitan et al. (2015) present the results of 
a comprehensive study of AM CVn systems over nearly 10 yr.
  They present outburst data on 32 systems with known orbital periods,
ranging from 5 to 65 min. A similar study with a slightly smaller
sample was presented by Ramsay et al. (2012).

 In Section 2 we discuss the DIM in the context of AM CVn accretion disks.
 In Section 3 we look at the recurrence time for superoutbursts, and in   
   section 4  the superoutburst duration.
   Section 5 summarizes our findings.

\section{The DIM in AM CVn systems}

\subsection{Background}

   A calculation of the vertical structure of geometrically
thin disks reveals
   the steady state physics underlying the limit
cycle behavior.
    the vertical structure as effective
temperature $T_{\rm eff}$  versus surface density
  $\Sigma$
    (Meyer \& Meyer-Hofmeister 1981, Cannizzo \& Wheeler 1984).
   One finds a hysteretic  relation 
 between midplane
  temperature  $T_{\rm mid}$ 
and surface density $\Sigma$, 
  or equivalently,
 between effective
  temperature  $T_{\rm eff}$ 
and $\Sigma$.
      Each radius $r$ and viscosity parameter $\alpha$ value has its own 
S--curve.   A parameterization of the 
   results of these calculations allows a
determination of the surface densities at the maxima
and minima in the S--curve, $\Sigma_{\rm max}(r, \alpha)$
and $\Sigma_{\rm min}(r, \alpha)$,
    as well as other physical quantities associated 
with these extrema, like midplane and effective temperature.
 One relates the local accretion rate to 
 $T_{\rm eff, max}$ and 
 $T_{\rm eff, min}$  using the standard Shakura \& Sunyaev (1973) relation
\begin{equation}
\sigma {T_{\rm eff}}^4 = {3 \over {8 \pi} } { G {\dot M} M_1 \over r^3}.  
\end{equation}

In the DIM,
    gas accumulates in quiescence and accretes onto
the central object in outburst.
   (e.g., Cannizzo 1993a, Lasota 2001 for reviews).
 The phases of quiescence and outburst
 are mediated by the action of heating and cooling fronts
that transverse the disk and bring about phase transitions
   between low and high states, consisting of neutral and ionized
gas, respectively.
    During quiescence, when the surface density $\Sigma(r)$ 
   at some  radius within the disk exceeds a critical value $\Sigma_{\rm max}(r)$,
      a transition  to the high state is initiated;
    during outburst, when $\Sigma(r)$ 
   drops below a different critical value $\Sigma_{\rm min}(r)$,
      a transition  to the low state is initiated.
Low$\rightarrow$high transitions can begin at any radius, 
whereas high$\rightarrow$low transitions begin at the outer disk edge.
  
  This situation comes about because of the following:
  In quiescence the disk is very non--steady so that mass accumulates
in the outer regions. The surface density distribution is bounded 
  by  $\Sigma_{\rm max}(r)$, which increases with radius. (Both 
  $\Sigma_{\rm max}(r)$ and $\Sigma_{\rm min}(r)$ scale close to linearly with $r$.)
In the outburst disk, however,  $\Sigma(r) \aprop   r^{-3/4}$.
    The accretion disk mass is conserved, 
  i.e., the disk mass accumulated by the end of quiescence $\Delta M_{\rm cold}$
      is the same as that in the hot disk $\Delta M_{\rm hot}$
 immediately  after the heating 
   transition has occurred.
   Therefore there must be a substantial redistribution of
        $\Sigma(r)$ --   from a profile $\aprop r$ in quiescence to $\aprop r^{-3/4}$ 
  in outburst.  Since  $\Sigma(r)$ is smallest at large radii in the outbursting
   disk, and since $\Sigma(r) < \Sigma_{\rm min}(r)$ is the condition for
  the cooling transition to begin, cooling fronts are always initiated in the 
     outer disk.

  One can define a ``maximum mass'' $\Delta M_{\rm max} \equiv
            \int 2\pi r dr \Sigma_{\rm max}(r)$  that the disk could possibly achieve
      during the accumulation phase, i.e., quiescence. 
 Obviously one cannot have
   $\Sigma(r) > \Sigma_{\rm max}(r)$  in  quiescence
   or else the instability would have already been
       triggered. In practice, time dependent calculations show that the true
  disk mass at the end of quiescence $\Delta M_{\rm cold}$ 
   is typically $\sim$$1/10-1/3$ of the ``maximum mass''
      (e.g., Cannizzo 1993b).  Therefore one may write
      $\Delta M_{\rm cold} = f \Delta M_{\rm max}$, 
       with $f \simeq 1/10 - 1/3$.

Levitan et al. (2015) restrict their attention to superoutbursts
     in their outburst statistics. These are analogous to long
outbursts in DNe above the period gap (van Paradijs 1983); 
    most of the mass accumulated during quiescence
accreted onto the primary before the disk shuts off.
   For normal, ``short'' outbursts,
    only a few percent of the stored gas accretes
  onto the central object:
         the thermal time scale of
   thin disk is short compared to the viscous time scale,
  and the cooling front that is launched from the outer  
  edge of the disk almost as soon as the disk  enters  into outburst
    traverses the disk and reverts it back to quiescence.
   For disks that have been ``filled'' to a higher level with 
respect to $\Delta M_{\rm max}$, the surface density
   in the outer disk can significantly exceed the critical surface
density $\Sigma_{\rm min}$.
     In order for the cooling front to begin,
      however,
       the outer surface density $\Sigma(r_{\rm outer})$
   must drop below 
    $\Sigma_{\rm min}(r_{\rm outer})$.
   Disks in this state generate much longer outbursts, with slower ``viscous''
plateaus, because the entire disk must remain in its high, completely ionized
state until enough mass has been lost onto the primary for the condition 
    $\Sigma(r_{\rm outer})  <  \Sigma_{\rm min}(r_{\rm outer})$  to be satisfied.

Various studies have investigated the DIM in AM CVn systems
  (Smak 1983a, Cannizzo 1984, Tsugawa \& Osaki 1997 = TO97,
 El--Khoury, \& Wickramasinghe 2000, Menou, Perna, \& Hernquist 2002,
 Lasota, Dubus, \& Kruk 2008 $=$ LDK,
   Kotko et al. 2012 $=$ KLDH).
These investigations generally consist of first calculating
  the steady state accretion disk structure by integrating
the vertical structure equations to parameterize the S--curve
relation between surface density $\Sigma(r,\alpha)$
and effective temperature $T_{\rm eff}(r, \alpha)$,
  and then using scalings for the steady state physics
   as input into a time dependent model to  calculate
 light curves.
  TO97 did not solve the full set of equations for the
vertical structure but rather prescribed a functional form
for the flux $F(z)$.         We restrict our 
 consideration of the scalings to 
   the two most recent studies LDK and KLDH,
   which integrate the 
complete set of structure equations and present a complete set of scalings
  for both the local minima and maxima in $\Sigma$.

As a model
    for the mass--losing secondaries in AM CVns,
Deloye et al. (2005) calculate pseudo--evolutionary sequences for
  donors with varying degrees of degeneracy.
  In their Figure 2 they calculate tracks in the ${\dot M}_T$--$P_{\rm orb}$
 plane for four isotherms based on central donor temperatures $T_c$
ranging from $10^4$ K to $10^7$ K, assuming a constant
   primary mass $0.6\msun$.
          They also indicate
   the upper and lower bounds of the instability strip for the DIM
 taken from TO97. For their two lowest  tracks, $T_c=10^4$ K and $10^6$ K,
    the instability strip spans roughly the correct 
              (i.e., observed) period range.
 Their tracks do not represent true evolutionary sequences
since  $T_c$ is taken to be constant along a track.

\subsection{Instability Criteria as Power Law Scalings}

The range of mass transfer from the secondary star
feeding into the outer disk ${\dot M}_T$ which allows for
unstable behavior, i.e., dwarf nova outbursts,
   is set by the local stability criteria
 at the inner and outer edge of the accretion disk.
 
KLDH calculate many S--curves for accretion disks relevant for
ultracompact binaries with no hydrogen, $X=0$.
     The basic finding is that the steady state scalings for the DIM
       are shifted to higher surface densities and temperatures.
  They present three sets of scalings for (i) $Y=1$, (ii) $Y=0.98$, $Z=0.02$,
  and (iii) $Y=0.96$, $Z=0.04$.
     Their results are given in terms of  $\alpha$, $r$, 
and $M_1$.
    The  range for unstable disk behavior is determined by
the S--curve for the inner and outer disk radii.
     If ${\dot M}_T >  {\dot M}_{T,2} \equiv{\dot M}_{\rm min, \ outer}$, 
the rate of accretion associated with the minimum in $\Sigma$
at $r_{\rm outer}$, the disk will be stable in the high, ionized state.
     If ${\dot M}_T < {\dot M}_{T,1} \equiv {\dot M}_{\rm max, \ inner}$, 
the rate of accretion associated with the maximum in $\Sigma$
at $r_{\rm inner}$, the disk will be stable in the low state.

 The KLDH scalings are not convenient; 
       for comparison with observations we must replace 
disk radius $r$ with orbital period $P_{\rm orb}$ in the KLDH
   scaling for ${\dot M}_{T,2}$, and with the primary radius
        in the ${\dot M}_{T,1}$ scaling.
  We follow TO97 in making these conversions:
  
(1)  For the outer scaling,
   ${\dot M}_{T,2}$, we relate the outer disk radius $r_{\rm outer}$
   to orbital period using Figure 5 from van Haaften et al. (2012).
     We identify the Roche lobe radius $R_{L1}$ with $r_{\rm outer}$
  and fit a simple power--law to $R_{L1}/a$
   as a function of the mass ratio $q=m_2/m_1$.\footnote{We
        use $M_1$ and $M_2$ to refer to the primary and secondary masses
in cgs units, and $m_1$ and $m_2$ to indicate $M/\msun$.}
         The  semi--major axis is $a$.
    Over the relevant range $0.01 \la q \la 0.03$ we fit
  $R_{L1}/a  \simeq 0.85 (q/0.01)^{-0.06}$.
    Based on the results of  van Haaften et al. (2012) 
      we adopt a fiducial secondary star $m_2(P_{\rm orb})$ relation 
\begin{equation}
   m_2 = 0.038  \left( P_{\rm orb} \over 1000 \ {\rm s} \right)^{-1.3}
\end{equation}
  (see their Sect. 2.3),
       corresponding to $R_2 \aprop M_2^{-0.18}$, which gives,
     using Kepler's law, 
\begin{equation}
  r_{o,10} = 1.18 m_1^{0.39} (1+q)^{1/3} 
       \left(P_{\rm orb} \over 
        1000 \ {\rm s} \right)^{0.74},
\end{equation}
   where $r_{o,10} = r_{\rm outer}/10^{10}$ cm.

(2)
  For  the inner scaling,
   ${\dot M}_{T,1}$, we adopt the standard inner boundary zero torque
condition (Shakura \& Sunyaev 1973) for which the
    maximum
   effective temperature  
 \begin{equation}
  T_{\rm eff}({\rm max}) = 0.488 \left(  3GM_{\rm wd} 
  {\dot M}  \over  8\pi \sigma R_{\rm wd}  \right)^{1/4}
 \end{equation}
  is reached at 49/36 times the inner edge, taken to be the primary radius.
   (The ``max'' in this equation refers to the local radial 
    maximum in $T_{\rm eff}$ in a steady state disk
  due to the inner boundary condition, not the 
            local maximum in $\Sigma$ in the S--curve.)
        Thus ${\dot M}_{T,1}$ is enhanced by a factor $(0.488)^{-4}$ and evaluated
  at 
$(49/36)R_{\rm wd}$.

  For the primary mass--radius relation, we fit a power law
    to Eggleton's scaling of a zero--temperature WD, i.e., 
  for which the ratio of atomic number to atomic weight $Z/A=1/2$ 
      (Rappaport et al. 1987,  see their eqn. [19], with $M_{\rm Ch} = 1.44\msun$). 
   Taking the scaling
 relevant for a He WD
    (van Haaften et al. 2012,  see their eqn. [25], with $M_p=5.66\times 10^{-4}\msun$),
    we tabulate values for $x=\log M_{\rm WD}$ and $y=\log R_{\rm WD}$
     over the  range of interest, 
           $0.5 < m_1  < 0.7$,
   and  fit a least squares power law $R_{\rm WD} =  
 10^{8.80}$ cm ${m_1}^{-0.62}$. 
    Over the fit range this 
          relation gives a maximum deviation  $<1.5$\% 
  from the Eggleton scaling.

  Applying these conversions to the scalings given in KLDH 
   (and ignoring the weak $\alpha$ dependencies)
           we obtain scalings 
     for ${\dot M}_{T,1}$ and ${\dot M}_{T,2}$ relevant for
their three compositions (i) $Y=1$, (ii) $Y=0.98$, $Z=0.02$, and (iii) $Y=0.96$, $Z=0.04$.
    Adopting the general forms 
\begin{flalign}
{\dot M}_{T,1} &= {\dot M}_{T,1,0} \ {m_1(P_2)}^{\epsilon_{1m1}} \\
{\dot M}_{T,2} &= {\dot M}_{T,2,0} \ {m_1(P_1)}^{\epsilon_{2m1}}  
                                 \  (1+q)^{\epsilon_q} \
   \left( P_{\rm orb} \over 1000 \ {\rm s} \right)^{\epsilon_p},
\end{flalign}
    we may now calculate the coefficients. 
 These are given in Table 1.


\begin{table}
\caption{KLDH-based coefficients for local extrema}
\centering
\begin{tabular}{ccccccc}
\hline
\hline
   {comp.} 
      & {${\dot M}_{T,1,0}$}
      & {${\dot M}_{T,2,0}$}
      & {$\epsilon_{1m1}$}
      & {$\epsilon_{2m1}$}
      & {$\epsilon_{q}$}
      & {$\epsilon_{p}$}  \cr
   {} 
      &({$\msunyr$})
      &({$\msunyr$})
      & {}
      & {}
      & {}
      & {}  \cr
\hline
  (i)  & $ 10^{-10.89} $
       & $  10^{-8.60} $
       & $-2.54$
       & $0.16$
       & $0.89$
       & $2.0$
   \cr
  (ii) & $  10^{-11.04} $
       & $  10^{-8.82} $ 
       & $ -2.50 $
       & $  0.16 $
       & $  0.89 $
       & $  1.99 $
   \cr
  (iii) & $  10^{-11.10} $
        & $  10^{-8.93}  $
        & $ -2.49 $
        & $  0.16 $
        & $  0.88 $
        & $  1.97 $  \cr
\hline
\end{tabular}
\label{t.R1}
\end{table}

       Levitan et al. (2015) presents a list of AM CVn 
   systems ordered 
by $P_{\rm orb}$, with outburst properties indicated.
The  rate of mass transfer ${\dot M}_T$ decreases sharply 
with $P_{\rm orb}$, so that the shortest $P_{\rm orb}$ systems
    have disks in permanent outburst,
those with intermediate $P_{\rm orb}$ exhibit outbursts,
and those with the longest $P_{\rm orb}$ have disks in permanent low states.
The dividing point between high state systems and outbursting systems
   lies at $P_1  \approx 20$ min.
The dividing point between 
   outbursting systems
  and 
  low state systems,
       $P_2$,  is not as straightforward. 
    A block of systems 
starting at $P_{\rm orb}=44.3$ min are listed as not showing outbursts\footnote{There
is also a system within the instability zone
                  at 35.2 min indicated as not having outbursts. 
      We disregard it on the assumption  it
    must be anomalous in some way, or  its outbursts may have been missed.},
but two systems among these do show 
             outbursts, at $P_{\rm orb}=47.3$ and 48.3 min.
   The expressions for ${\dot M}_{T,1}$  indicate a steep inverse scaling
 with $m_1$, therefore it seems probable that these systems
   have somewhat high primary masses. Therefore we set $P_2=44$ min.
        Furthermore  we adopt $m_1(P_1) = m_1(P_2)=0.6$.
    We note that the $m_1$ values for AM CVn systems
      are not well-constrained observationally from dynamical measurements.

We may now use the observed instability strip for AM CVn systems
  to constrain the secondary mass transfer rate ${\dot M}_T$.
  Let us assume a power law 
  $ {\dot M}_T = A \left( P_{\rm orb}  /  1000 \ {\rm s} \right)^n$.
     For simplicity we adopt the convention that ${\dot M}_T > 0$.
    Since we expect  $ {\dot M}_T$ to decrease with orbital period,
   we     set ${\dot M}_{T,1} ={\dot M}_T(P_2)$ and 
              ${\dot M}_{T,2} ={\dot M}_T(P_1)$, 
    which gives
      two equations in two unknowns.
   We may solve for the normalization constant $A$ and dependence
on orbital period $n=d\ln {\dot M}_T/d\ln P_{\rm orb}$.
   This yields the general solution 
\begin{flalign}
 n&= 
 {\log[ ({\dot M}_{T,2,0} / {\dot M}_{T,1,0}) 
    \ g \ 
         (P_1/1000 \ {\rm s})^{\epsilon_p} 
       ]
    \over \log(P_1/P_2)}  \\
 A&= {\dot M}_{T,1,0}  \
    \ {m_1(P_2)}^{\epsilon_{1m1}}
    \  \left( P_2 \over 1000 \ {\rm s} \right)^{-n},
\end{flalign}
  where 
\begin{equation}
 g = 
      (1+q)^{\epsilon_q} 
     m_1^{\epsilon_{2m1} - \epsilon_{1m1}}. 
\end{equation}

 Table 2 gives the values for $A$ and $n$
      for the three KLDH compositions, adopting 
    $P_1=20$ min, $P_2=$ 44 min,
        and  $m_1=0.6\pm0.05$.
         The magnitude of the 
   putative assigned error on $m_1$
    was propagated through to $n$ and $A$ in order to indicate the strength of
the dependency.
    The main uncertainty entering into $n$ and $A$ 
  is the assumed primary mass. 
   For instance, taking $m_1=1$ 
    would give $n \simeq -7$, a much 
           steeper ${\dot M}_T (P_{\rm orb})$ relation.

 \begin{table}
\centering 
 \caption{{Coefficients for 
   $ {\dot M}_T = A ( P_{\rm orb} / 1000 \ {\rm s} )^n$}
        with $ m_1=0.6\pm0.05$  }
\tablewidth{0pt}
 \begin{tabular}{ccc}
\hline
\hline
 composition   & 
$A(\msunyr)$  &
$n$
\cr
\hline
  (i)  & $ 8.82\pm 0.6  \times 10^{-9}  $
       & $ -5.38\pm 0.3 $
   \cr
  (ii) & $ 5.27\pm 0.4  \times 10^{-9}  $
       & $ -5.23\pm 0.3 $
   \cr
  (iii) & $  3.88\pm 0.3  \times 10^{-9}  $
       &  $ -5.06\pm 0.3 $
   \cr
\hline
 \end{tabular}
\label{t.R2}
 \end{table}

Since precise abundances
   for AM CVns as a group are not known,
    we adopt a 
      representative
     fiducial scaling ${\dot M}_T \simeq 5\times 10^{-9} \msunyr$
$(P_{\rm orb}/1000 \ {\rm s})^{-5.2}$,
  relevant for
   the 
   middle range of scalings given in KLDH.

\subsection{Considerations from AM CVn Binary Evolution}

How does this $\dot{M}_T$ law compare to theoretical expectations? As
mentioned above, in AM CVn systems the driving of the mass transfer is
most likely the angular momentum loss due to GWR. For that case one can
derive the expected scaling of mass transfer rate with orbital period
as is done in Warner (1995b). We begin by expressing the mass transfer rate
in   terms of the 
    angular momentum loss 
  \begin{equation}
\frac{\dot{M_2}}{M_2} \propto \left(\frac{\dot{J}}{J_{\rm
orb}}\right)_{\rm GWR} \propto \frac{M_1 M_2 (M_1 + M_2)}{a^4}
\end{equation}
   (e.g., 
       see Savonije  et al. 1986;
  Marsh et al. 2004;
    eqn. [6]
of van Haaften et al 2012),
        where ${\dot M}_2 = -{\dot M}_T$.
Assuming $M_1$ remains constant and $M_2 \ll M_1$ and using Kepler's law
to replace orbital separation with orbital period,
         we find
\begin{equation}
\frac{\dot{M_2}}{M_2} \propto M_2 P_{\rm orb}^{-8/3}.
\end{equation}
We then can use the fact that the size of the Roche lobe for
given period depends only very weakly on the mass of the accretor (the
well known period--mean density relation) to find how $P_{\rm orb}$
scales with $M_2$, given the mass--radius relation for the donor
$R_2 \propto M_2^\zeta$. Hence
\begin{equation}
P_{\rm orb} \propto \left(\frac{R_2^3}{M_2}\right)^{1/2}  \propto M_2^{(3 \zeta
-1)/2}
\end{equation} 
     and therefore 
\begin{equation}
\dot{M_2} \propto M_2^2 P_{\rm orb}^{-8/3} \propto P_{\rm
orb}^{  4/(3 \zeta - 1) - 8/3 }.
\end{equation}
For $\zeta = -1/3$ we find an
  exponent\footnote{Note that
there is an exponent ``$-1$'' missing in eqn. (4) of Warner (1995b); for
$\zeta = -1/3$ the donor mass and orbital period scale inversely, not
linearly.} 
      $n=-14/3$.
       In van Haaften et al. (2012) the values of $\zeta$ are plotted
in their Figure~2.    For low masses  $\zeta > -1/3$,
yielding a larger absolute value of the exponent. Indeed, the fit they make to
the dependence of mass transfer rate 
  on orbital period (their Appendix
A) gives $n=-5.32$ for a 1.4 $\msun$ accretor. 
            Warner (1995b) uses $\zeta
= -0.19$, based on Savonije et al. (1986), and finds $n=-5.21$. 
       We  conclude
that the exponent for the scaling we infer
  from the DIM based on the observed instability strip for AM CVns
                    is in good agreement with 
  expectations from 
    stellar structure
        if we adopt $m_1\simeq0.6$.

\section{Recurrence Time for Outbursts}

For 11 of the 32 
    AM CVn systems
with known orbital periods in their study, Levitan et al (2015)
     have enough coverage to make quantitative statements
     about their outburst properties.  They
find a relation for the recurrence time for outbursts
\begin{equation}
t_{\rm recur} = 1.46 \ {\rm d} \ \left(P_{\rm orb} \over 1000 \ {\rm s}\right)^{7.35}
    + 24.7 \ {\rm d}.
\end{equation}
  They only include superoutbursts. 
   Their relation (see their Figure 12a)
   is much tighter  
than the comparable plot for DNe above the minimum period
(e.g.,    see  Figures 2--4 of CSW;
      Figure 11 of Patterson 2011).

What is the expectation from DIM for the recurrence time and is it
consistent with this scaling?

   CSW formulated an analytical expression for
  $t_{\rm recur}$. Their full complexity is not needed, and indeed 
one can take a rather simple approach. 
   Regardless of whether one subscribes to the thermal--tidal
 instability for superoutbursts or the plain DIM (Osaki \& Kato 2013),
  the normal outbursts in a system exhibiting both normal outbursts
and superoutbursts represent a tiny fraction of the mass budget.
    Therefore a good approximation is that superoutbursts
are the only outbursts, and that during a superoutburst essentially
all the mass stored in the cold state is accreted.
   The recurrence time is then simply
\begin{equation}
  t_{\rm recur} =  {  \Delta M_{\rm cold} \over {\dot M}_T} 
          =  { f \Delta M_{\rm max} \over {\dot M}_T} =
  {f \int 2\pi r dr \Sigma_{\rm max}(r,\alpha) \over {\dot M}_T},
\end{equation}
 where $\Delta M_{\rm max}$ is the maximum mass
   that could be  stored in quiescence
and $f$ is the fraction  the disk  is filled,
relative to this maximum.

  For specificity we adopt the middle of the three compositions
     considered by KLDH.
Their $\Sigma_{\rm max}$ for $Y=0.98$, $Z=0.02$ is
\begin{equation}
\Sigma_{\rm max} = 612 \ {\rm g} \ {\rm cm}^{-2} \ 
     {\alpha_{c-1}}^{-0.82} \
    {r_{10}}^{1.10} \ 
    {m_1}^{-0.37},
\end{equation}
  where $\alpha_{c-1}$ is the value of $\alpha$ in the
low state of the disk,  $\alpha_{\rm cold}$, normalized to 0.1.
  Hence
\begin{multline}
\Delta M_{\rm max} = 1.04\times 10^{-9} \ \msun \ 
     {\alpha_{c-1}}^{-0.82} \
    {m_1}^{0.85} \\
   (1+q)^{1.03} \
    \left(P_{\rm orb} \over 1000 \ {\rm s}\right)^{2.31}.
\end{multline}
   Evaluating 
     the recurrence time gives
\begin{flalign}
  t_{\rm recur} 
    & = 7.59 \ {\rm d} \
           f_{-1}  \
     {\alpha_{c-1}}^{-0.82} \
    {m_1}^{0.85} \
   (1+q)^{1.03} \
   \left( P_{\rm orb} \over 1000 \ {\rm s} \right)^{7.51} 
    \\
     &  = 4.92 \ {\rm d} \
   \left( P_{\rm orb} \over 1000 \ {\rm s} \right)^{7.51}
\end{flalign}
    for $m_1=0.6$, where $f_{-1} = f/0.1 = \alpha_{c-1} = 1$.
   Thus for a mass transfer rate with $n=
 d\ln  {\dot M}_T/d\ln  P_{\rm orb} = -5.2$,  the simplest recurrence
  time scaling gives $t_{\rm recur} \aprop  {P_{\rm orb}}^{7.5}$,
  which is close to the observed relation.

\section{Duration Time for Outbursts}

For
the AM CVn  superoutburst durations    
Levitan et al. (2015) find 
\begin{equation}
   t_{\rm dur} = 0.89 \ {\rm d} 
   \ \left( P \over 1000 \ {\rm s} \right)^{4.54} +10.6 \ {\rm d}.
\end{equation}
   This relation appears to 
  be less reliable than their $t_{\rm recur}(P_{\rm orb})$ relation:
  for the four longest period systems only upper limits
   are given. In fact, 
  for the other six systems, 
 those with  $22$ min $< P_{\rm orb} < 29$ min,
      their data are
   consistent with 
    $t_{\rm dur}(P_{\rm orb})$ 
   being constant  with orbital period.
 We note that the four outburst durations given as upper limits in 
    Table 2 of Levitan et al. are plotted as actual values in 
their Figure 12, panel 3, and enter into their power law 
   fitting.
    As noted in their Appendix A, 
        for the four systems with only upper limits 
        $(UL)_i$ on $t_{{\rm dur} \ i}$
        they arbitrarily take $t_{{\rm dur} \ i} = 0.75(UL)_i$, 
   with error $0.25(UL)_i$.

What would one expect from theory?

There are several ways to construct 
     an outburst duration timescale
  for superoutbursts.
     During a long outburst with a ``viscous plateau'',
  the cooling front cannot propagate due to excess surface
  density at the outer disk edge, relative to $\Sigma_{\rm min}$,
    therefore the only option is for accretion onto the primary
  so that $\Sigma(r_{\rm outer})$ is gradually reduced.
    Therefore it is reasonable to consider the viscous time 
  scale in the outer disk in the hot state
         as an approximation to the superoutburst duration.

   We follow the method given in King \& Pringle (2009).
     They estimate a local peak  accretion rate during outburst
   as $\simeq 2 \Delta M_{\rm hot}/t_{\rm visc}$,
    where $\Delta M_{\rm  hot}$ is the mass of the hot disk.
       The outbursting disk mass is determined by that stored
in quiescence,  $\Delta M_{\rm  hot}= \Delta M_{\rm cold}$.
 This eliminates ${\dot M}$ from the 
  standard equations given in  Frank, King, \& Raine (2002),
   which 
   relate   the locally defined viscous time to $\alpha$, $r$, and
  ${\dot M}$. 
         Using the outer disk edge for $r$ and relating  
      it to orbital period as previously gives 
\begin{flalign}
t_{\rm visc} &= 15.1 \ {\rm d} 
                 \ \alpha_{h-1}^{-0.8}
                 \ m_1^{0.34}
                 \ (1+q)^{0.16} 
                 \ \left( P_{\rm orb} \over 
      1000 \ {\rm s}\right)^{0.36}
\end{flalign}
 where $\alpha_{h-1}=\alpha_{\rm hot}/0.1$.
 
This is much flatter than the Levitan et al. relation
    for  $t_{\rm dur}(P_{\rm orb})$, but their fitted relation
  is dominated  
   by  including upper limits for  $t_{\rm dur}$
 (for
   systems with $P_{\rm orb} > 30$ min)
   as part of their fit,
     which makes a direct
comparison with theory problematic.

\section{Discussion and Conclusion}

Previous workers have examined the stability properties
  of AM CVn systems in the context of expectations from the DIM,
  e.g.,  
  TO97 (see their Fig. 4),
  Nelemans (2005, see his Figs. 1 and 6),
   LDK (see their Fig. 3),
  and 
  KLDH (see their Fig. 3).
      The last work, KLDH, has provided the most complete study to date.
     Although 
  they do not provide explicit formulae between the
   critical mass transfer rates and orbital period as in TO97,
       KLDH plot them in their Fig. 3, as do the aforementioned studies.
  Our main difference with KLDH is that for ${\dot M}_{T,1}$ 
     we also take into
   account the zero torque boundary condition at the inner edge of
  the disk (following TO97), 
              in addition to just 
       considering the primary mass. 
          However, 
   ${\dot M}_{T,1}$ depends steeply on the primary mass ($\aprop m_1^{-2.5}$), 
        and this uncertainty
will likely dominate that associated with 
          the  boundary condition refinement.
      KLDH  plot in their Fig. 3 the expected secular  ${\dot M}_T$
 versus $P_{\rm orb}$ from stellar evolutionary models,  
         and find consistency with the DIM in terms of stability of 
   observed systems.  Thus the difference between our study and KLDH
  is one of perspective: KLDH combine observations with evolutionary models 
       to show that AM CVn outbursts can be explained by the DIM,
   whereas we assume that the DIM is correct in order to derive
       ${\dot M}_T(P_{\rm orb})$, and then explore the ramifications
  of the derived law vis a vis not only   ${\dot M}_T(P_{\rm orb})$
   predicted from evolutionary models, but also the resultant outburst
 properties versus  $P_{\rm orb}$.

   In summary,
we apply the DIM model to the recent results on AM CVn
   systems obtained 
 by Levitan et al. (2015), using scalings for the helium--rich accretion 
disks in such systems taken from Kotko et al. (2012).
  The orbital periods defining the edges of the 
  instability strip $P_1$ and $P_2$ permit us to infer
     a mean secondary mass transfer rate ${\dot M}_T
  \simeq 5 \times 10^{-9} \msunyr (P_{\rm orb}/1000 \ {\rm s})^{-5.2}$.
   Our finding of a      steep inverse dependence $n\simeq -5.2$
   is consistent with theoretical expectations,
   but our result is dependent on taking $m_1\simeq0.6$; 
   higher $m_1$ steepens the relation. 
         Treating $m_1$ as a variable and all other parameters
               on the right hand side of equation (7) as constant
     we may write $n \approx -5.23 - 7.75 \log(m_1/0.6)$
     for $Y = 0.98$, $Z = 0.02$. The largest 
        uncertainty in $n$ enters via $m_1$.
 
    We  emphasize that the power-law form for $P_{\rm orb}$ 
 is not an outcome of our analysis but an assumption, valid in our
application only over $20$ min $\la P_{\rm orb} \la 45$ min.
      However, our inferred  $P_{\rm orb}$ values at  $20$ and $45$ min
  are in line with those estimated for stable systems 
 with similar orbital periods (KLDH, see their Fig. 3).
   We note that the precise values of  $P_{\rm orb}$
         defining the edge of the instability strip, $P_1$ and $P_2$,
   also affect $n$ and $A$, although not as strongly as $m_1$.   Lastly, 
   $n$ and $A$ are only weakly dependent on 
   the   composition of the gas 
   and 
    the viscosity parameter $\alpha$; these two factors enter via 
         ${\dot M}_{T,2,0}/{\dot M}_{T,1,0}$ (see eqn. [7]).  
   This ratio is relatively insensitive to composition and 
   virtually independent of $\alpha$. 
    The full dependencies of $n$ and $A$ on  all input parameters are given in eqs. (7)-(9).

 The simplest possible theoretical expression
   for the recurrence time from the DIM gives $t_{\rm recur} \propto
  {r_{\rm outer}^{3.1}}/{\dot M}_T  \propto 
   {P_{\rm orb}}^{2.31} {{\dot M}_T}^{-1}$,
  so that  $(d\ln  t_{\rm recur}/d\ln  P_{\rm orb}) 
   =  2.31 - (-5.2) \simeq 7.5$,  close to the value 7.35 found by Levitan et al. (2015).
      However, a larger assumed value of $m_1$ in our model would increase the
exponent. 
      Thus, if the DIM is relevant for AM CVn outbursts, the primaries
       must have masses $m_1 \simeq 0.6$.
   The Levitan et al. constraint on the outburst duration $\propto P_{\rm orb}^{4.54}$
appears to be dominated by incompleteness for the upper half of their range
in orbital period. Our theoretical prediction is that a more complete time
sampling of AM CVn outbursts, especially at longer orbital period, will
  ultimately reveal a much flatter law $t_{\rm dur} \aprop P_{\rm orb}^{0.4}$.

\smallskip
\smallskip

We thank Joe Patterson for organizing a small CV
   workshop at Columbia University in Fall of 2014 
   which provided a stimulating environment
for discussion. 
    We also thank the anonymous referee whose comments improved the paper.

\vfil\eject

\def\mnras{MNRAS}
\def\apj{ApJ}
\def\apjs{ApJS}
\def\apjl{ApJL}
\def\aj{AJ}
\def\araa{ARA\&A}
\def\aap{A\&A}




\begin{references}

\reference{}
Cannizzo, J.~K. 1984, Nature, 311, 443

%
\reference{}
Cannizzo, J.~K. 1993a, in Accretion Disks in
       Compact Stellar Systems,
   ed. J. C. Wheeler (Singapore: World Scientific), 6
%

\reference{}
Cannizzo, J.~K. 1993b, ApJ, 419, 318
%

\reference{}
Cannizzo, J.~K., Shafter, A.~W., \& Wheeler, J.~C. 1988, ApJ, 333, 227 (CSW)

\reference{}
Cannizzo, J.~K., \& Wheeler, J.~C. 1984, ApJS, 55, 367

\reference{}
Deloye, C.~J., Bildsten, L., \& Nelemans, G.
      2005, ApJ, 624, 934

\reference{}
El-Khoury, W., \& Wickramasinghe, D. 2000, A\&A, 358, 154

\reference{}
Faulkner, J., Flannery, B.~P., \& Warner, B. 
       1972, ApJ, 175, L79

\reference{}
Frank, J., King, A.~R., \& Raine, D.~J. 2002,
    Accretion Power in Astrophysics, 3rd ed.
    (Cambridge: Cambridge Univ. Press)

\reference{}
Groot, P.~J., Nelemans, G., Steeghs, D., 
        \& Marsh, T.~R. 2001, ApJ, 558, L123

\reference{}
King, A.~R., \& Pringle, J.~E. 2009, MNRAS, 397, L51


\reference{}
Knigge, C.,  Baraffe, I., \& Patterson, J.
    2011, ApJS, 194, 28  (KBP)

\reference{}
Kotko, I., Lasota, J.-P., Dubus, G., 
   \& Hameury, J.-M. 2012, A\&A, 544, A13 (KLDH)

\reference{}
Lasota, J.-P. 2001, New Astron. Rev., 45, 449
%

\reference{}
Lasota, J.-P., Dubus, G., \& Kruk, K. 2008, A\&A, 486, 523 (LDK)

\reference{}
Levitan, D., Groot, P.~J.,  Prince, T.~A., Kulkarni, S.~R.,
    Laher, R., Ofek, E.~O., Sesar, B., \& Surace, J.
   2015, MNRAS, 446, 391


\reference{}
Marsh, T.~R., Nelemans, G., \& Steeghs, D. 2004, 
     MNRAS, 350, 113


\reference{}
Menou, K., Perna, R., \& Hernquist, L. 2002, ApJ, 564, L81

\reference{}
Meyer, F. \& Meyer-Hofmeister, E. 1981, A\&A, 104, L10


\reference{}
Nather, R.~E., Robinson, E.~L., \& Stover, R.~J. 1981,
     ApJ, 244, 269


\reference{}
Nelemans, G. 2005, in The Astrophysics of Cataclysmic Variables
     and Related Objects, eds. J.-M. Hameury, \& J.-P. Lasota,
      Proc. ASP Conf., 330, 27 


\reference{}
Osaki, Y., \& Kato, T. 2013, PASJ, 65, 50

\reference{}
  Patterson, J.~P. 2011, MNRAS, 411, 2695

\reference{}
  Patterson, J., Halpern, J., \& Shambrook, A. 
       1993, ApJ, 419, 803


\reference{}
Ramsay, G., Barclay, T., Steeghs, D., Wheatley, P.~J., 
      Hakala, P., Kotko, I., \& Rosen, S. 2012, MNRAS, 419, 2836

\reference{}
Rappaport, S., Ma, C.~P., Joss, P.~C., \& Nelson, L.~A. 1987, ApJ, 322, 842

\reference{}
Savonije, G.~J., de Kool, M., \& van den Heuvel, E.~P.~J. 1986,
              A\&A, 155, 51

\reference{}
Shakura, N.~I., \& Sunyaev, R.~A. 1973, A\&A, 24, 337

\reference{}
Smak, J. 1983a, Acta Astron., 33, 333

\reference{}
Smak, J. 1983b, ApJ, 272, 234

\reference{}
Smak, J. 1984, Acta Astr., 34, 161

\reference{}
Solheim, J.-E. 2010, PASP, 122, 1133

\reference{}
Tsugawa, M., \& Osaki, Y.
       1997, PASJ, 49, 75 (TO97)

\reference{}
van Haaften, L.~M.,  Nelemans, G., Voss, R., Wood, M.~A., \& Kuijpers, J. 
           2012, A\&A, 537, A104

\reference{}
van Paradijs, J. 1983, A\&A, 125, L16

\reference{}
Warner, B. 1995a, 
   Cataclysmic Variable Stars (Cambridge: Cambridge Univ. Press)
  
\reference{}
Warner, B. 1995b, Ap\&SS, 225, 249  

\reference{}
Warner, B., \& Robinson, E.~L. 1972, MNRAS, 159, 101


\end{references}
\end{document}